\documentclass[preprint,showpacs,prb,preprintnumbers,amsmath,amssymb,floatfix]{revtex4}
\usepackage{amsmath}
\usepackage{graphicx}% Include figure files
\usepackage{dcolumn}% Align table columns on decimal point
\usepackage{bm}
\usepackage{subfigure}
\usepackage[utf8]{inputenc}
\usepackage{color}
               
\begin{document}

\title{Explicit spatial description of fluid inclusions in porous
  matrices in terms of an inhomogeneous integral equation} %Title of paper

\author{E. Lomba}
\email{enrique.lomba@csic.es}
\author{Cecilia Bores}
\affiliation{Instituto de Química Física Rocasolano, CSIC, Serrano
  119, E-28006 Madrid, Spain}
\author{Gerhard Kahl}
\affiliation{Institut für Theoretische Physik, Technische Universität
  Wien and Center for Computational Materials Science (CMS), Wiedner Hauptstraße 8-10, A-1040 Wien, Austria}

%\homepage[]{Your web page}
%\thanks{}

\date{\today}
\begin{abstract}
We study the fluid inclusion of both Lennard-Jones particles and particles with competing
interaction ranges --short range attractive and long range repulsive (SALR)-- in a disordered
porous medium constructed as a controlled pore glass in two
dimensions. With the aid of  a full two-dimensional 
Ornstein-Zernike approach, complemented by a Replica Ornstein-Zernike
integral equation,  we explicitly obtain the spatial density distribution of the
fluid adsorbed in the porous matrix and a good approximation for the
average fluid-matrix correlations. The results illustrate the
remarkable differences between the adsorbed Lennard-Jones (LJ) and SALR systems. In the
latter instance, particles tend to aggregate in clusters which occupy
pockets and bays in the porous structure, whereas the LJ fluid
uniformly wets the porous walls. A comparison
with Molecular Dynamics simulations shows that the two-dimensional Ornstein-Zernike
approach with a Hypernetted Chain closure together with a sensible approximation
for the fluid-fluid correlations
can provide an accurate picture of the spatial distribution of 
adsorbed fluids for a given configuration of porous material.

% insert abstract here
\end{abstract}

\pacs{68.43.-h, 68.43.Hn, 61.43.Bn, 61.43.Gt}% insert suggested PACS numbers in braces on next line

\maketitle %\maketitle must follow title, authors, abstract and \pacs

\section{Introduction}
The study of fluid inclusions and/or adsorption in a porous matrix from an atomistic
standpoint is essential to get a better understanding of key
technological issues such as molecular sieving, heterogeneous
catalysis or gas storage. From the theoretical perspective, the advent
of the Replica Ornstein Zernike (ROZ) approach in the early
nineties\cite{JStatPhys_1988_51_537,JCP_1992_97_4573,PRE_1993_48_233}
provided a powerful alternative to direct molecular simulation for 
the description of fluid inclusions in disordered porous systems.
Since then, the ROZ approximation has been much exploited to describe templated
\cite{PRE_1999_60_R25,JCP_2000_112_3006,JCP_2005_123_164706,JCP_2008_128_044707}
and sponge-like
materials\cite{JCP_2006_125_244703,JCP_2007_127_144701}, and a large
variety of inclusions, such as simple binary mixtures\cite{PRE_2000_61_5330}
illustrating their phase behavior\cite{Scholl2001}, colloid/polymer
mixtures\cite{JPCM_2008_20_115101}, electrolytes
\cite{JPCB_1999_103_5369,PhysA_2002_314_156,JPCB_2004_108_1046,JCP_2010_132_104705},
and associating fluids\cite{JML_1998_76_183,JML_2004_112_71}. This
approach yields average thermodynamic properties, fluid-fluid, and
fluid-matrix correlations, but if one is interested in the explicit spatial
distribution of the fluid/adsorbate for a given configuration of the
matrix an alternative approach is needed, aside from resorting to
molecular simulation. This structural information is
particularly important when dealing with functionalized adsorbents
(see e.g. Wood et al.\cite{Wood2012} and references therein) both for
gas storage or catalysis, since the particular location of adsorbents
or reactants within the substrate is crucial to evaluate whether the
adsorbent material has the desired properties.

This is in principle a challenging theoretical
problem, in which we have to solve the statistical mechanics
of a fluid in the presence of a highly
non-uniform (and topologically disordered) external field stemming from the adsorbent
matrix. Interestingly, an avenue to tackle this problem was opened two
decades ago by Beglov and
Roux\cite{JCP_103_360}, who explored the ability of the Hypernetted Chain
equation (HNC) to describe the solvation of solutes with arbitrary
geometry when treated explicitly in three dimensions. Related
approximations have 
been exploited with great success to study the physics of solvation of
complex molecules\cite{Beglov1996,JPCB_104_796,Kovalenko1998}, including 
proteins\cite{Perkyns2010,Yamazaki2011}. It just turns out that 
Beglov and Roux\cite{JCP_103_360}, also explored the possibility of applying their
approach to confined fluids in order  to analyze the density profile of a monoatomic
Lennard-Jones (LJ) fluid adsorbed in a simplistic model of
zeolite. The approach proved to be relatively successful,
despite the use of a crude approximation consisting in the replacement
of the fluid-fluid direct correlation under confinement by its bulk counterpart. 

The purpose of this paper is to study in full detail the capabilities of
Beglov and Roux's integral equation approach in a model system that illustrates  the effects
of confinement on the spatial distribution of adsorbates inside a given
topological configuration of the confining matrix. At the same
time, we intend to 
provide a somewhat more elaborate approximation for the fluid-fluid
correlations under confinement. To that aim we have analyzed the
behavior of a two dimensional fluid with competing interaction ranges
whose particles tend to cluster at low 
temperatures. Our model is a soft core version of the
``short-range attractive and  long-range repulsive'' (SALR) potential first
proposed by Sear and coworkers\cite{Sear1999}, and analyzed in detail by 
Imperio and Reatto\cite{JPCM_2004_16_S3769,Imperio2006}. The 
clustering properties under disordered confinement of this hard core model have
also recently been studied by Schwanzer and Kahl\cite{Schwanzer2011}. In our case, the matrix-fluid
interactions are purely repulsive and soft. For the sake of comparison,
we have also considered the
inclusion of a two dimensional LJ fluid which interacts with matrix
particles via a LJ potential as well. 

In order to create a disordered matrix with a
relatively large porosity, we have used the templating approach
characteristic of the fabrication of controlled porous
glasses\cite{Lang_1998_14_2097}, using as precursor a mixture of
non-additive hard disks. Once frozen for a total density
slightly below the demixing critical density\cite{Buhot2005}, one of
the components of the mixture is removed, together with all disconnected
clusters of particles of the remaining
component\cite{Lang_1998_14_2097}. In this way, we have generated a
system with a high degree of topological disorder but with enough free
space to enable the formation of clusters and/or the condensation of
the Lennard-Jones fluid. 

Our system of interest will be then an inclusion of a thermally equilibrated
fluid into one particular disordered matrix configuration. Now,  with the purpose of 
providing a sensible approximation for the fluid-fluid direct correlation under confinement required within the formalism, we have resorted to a ROZ equation
with an HNC closure for a templated
matrix\cite{JCP_2000_112_3006,JCP_2010_132_104705} (ROZ-HNC). This approach
furnishes fluid-fluid correlations 
averaged over matrix disorder, and these will be seen to represent a
better approximation for our adsorbed fluid than the corresponding
bulk counterparts. Note that our
system is self-averaging, and hence the thermal average of the fluid
correlations for a selected large matrix configuration can adequately be
approximated by its average over disorder. We will see that the two-dimensional
Hypernetted-Chain approach (HNC-2D) complemented by the ROZ-HNC equation
provides an excellent description 
of the spatial distribution of the adsorbed fluid. 

The rest of the paper is sketched as follows. In the next section we
describe in detail our model. The key elements of the HNC-2D approach
are presented in Section III together with a summary of the main
equations of the ROZ-HNC theory. Our most significant results and
future prospects are collected in Section IV.

\section{The model}
Our system is formed by a disordered matrix and an annealed
fluid.  Strictly speaking, we would be dealing with a three component
system consisting in a mixture of non-additive hard disks (components
$\alpha$ and $\beta$, being $\beta$ the template) and the fluid. Then components $\alpha$ and $\beta$
are quenched, and component $\beta$ is removed. In addition, those $\alpha$ particles that remain disconnected after the removal of the $\beta$ particles are also removed. Connectivity of the matrix is analyzed by identifying
clusters of matrix particles, and two particles are considered to be
linked in a cluster if their separation is smaller than two particle
diameters. All disconnected clusters formed by less than ten particles
are removed. This procedure attempts to roughly mimic the template and
some loose matrix particles being washed away from the porous matrix. 

 The fluid can be trapped in disconnected
cavities, and in this regard, the problem is somewhat different from
a process of fluid  adsorption. In this latter instance particles diffuse only
through percolating pores. However, from a practical point of view,  fluid chemical potentials, excess internal energies per
particle and averaged correlation functions, in our case are found to
be very similar, whether the presence of fluid particles inside
disconnected cavities is allowed or not.
Consequently, we will use the terms inclusion and adsorption
indistinctly. 

%%%

As mentioned, our porous matrix is built by quenching a symmetric binary system
of non-additive hard disks interacting through a potential of the form
\begin{equation}
\beta u_{0_i0_j}(r) = \left\{\begin{matrix}
\infty & \mbox{if}\; r < \sigma(1+\Delta(1-\delta_{ij})) \\
0 & \mbox{otherwise}
\end{matrix}
\right.
\label{vmm}
\end{equation}
where, the subscript 0 will denote the matrix particles, $i,j$ are
$\alpha$ or $\beta$ and $\delta_{ij}$ is a Kronecker's $\delta$. We have considered a non-additivity parameter
$\Delta=0.2$, and the configurations of the matrix have been generated
for  total densities $(\rho_{0_\alpha}+\rho_{0_\beta})\sigma^2=0.632$,
and $(\rho_{0_\alpha}+\rho_{0_\beta})\sigma^2=0.675$, 
which lie somewhat below the critical demixing density which we have
calculated\cite{Bores2014}, $\rho_c\sigma^2=0.684(1)$. Particles of type
$\beta$ are removed, and then disconnected $\alpha$ particles are
removed as well, by which our final 
matrix densities will be $\rho_0\sigma^2=0.314$, and $\rho_0\sigma^2=0.324$ with
$\rho_0=\rho_{0_\alpha}\approx \rho_{0_\beta}$. These densities
correspond to the specific matrix configurations for which our
calculations are carried out. As usual, we define the porosity of the
matrix as the
ratio of area available for the insertion of a test fluid particle in
an empty matrix configuration with respect to the total matrix sample
area. With this definition in mind, we find that even if
our two matrix densities are quite close, the  porosity  in the latter
instance is appreciably larger (47.7\%  for $\rho_0\sigma^2=0.324$ vs
41.5\% for $\rho_0\sigma^2=0.314$). This results from
the density of the precursor non-additive hard 
disk mixture being closer to the consolute point, by which one-component
clusters are larger and tend to span the whole simulation box. Note also that for
arriving at $\rho_0\sigma^2=0.324$ a substantial number of disconnected $\alpha$
particles had to be  removed (the original value for the density of the $\alpha$-particles we started from was $\rho_{0_\alpha}\sigma^2 = 0.338$);
 this was not the case for  $\rho_0\sigma^2=0.314$ (the original
value for the density, $\rho_{0_\alpha}\sigma^2 = 0.316$, is quite close to the actual one).

As to the fluid inclusion, we will first consider a system with competing
interactions (SALR)  of the type studied by Imperio and Reatto\cite{JPCM_2004_16_S3769}
\begin{equation}
u_{11}(r) =
-\frac{\varepsilon\sigma^2}{R_a^2}\exp(-r/R_a)+\frac{\varepsilon_r\sigma^2}{R_r^2}\exp(-r/R_r)
+u_{sr}(r).
\end{equation}
with $\varepsilon_r=\varepsilon$, $R_r=2R_a=2\sigma$. Here and in what
follows, the
subscript 1 denotes fluid particles. For practical
purposes, we have used a soft highly repulsive 
interaction of the form
\begin{equation}
u_{sr}=\varepsilon\left(\frac{\sigma(1-\delta)}{r}\right)^{20}
\label{ur}
\end{equation}
where we have set $\delta=0.01$. This choice of potential reproduces the
same contribution of the SALR potential to the internal energy and
the same pressure as the hard disk model of
Ref.~\onlinecite{JPCM_2004_16_S3769}, for temperatures above
$k_BT/\varepsilon>1$, where $k_B$ is Boltzmann's constant as usual and
$T$ the absolute temperature.  The matrix fluid interaction is  given by
\begin{equation}
u_{01}(r) = u_{sr}(r)
\label{u01}.
\end{equation}
and Eq.~(\ref{ur}). 

Together with the SALR potential, we have also studied a system of
LJ disks for which
\begin{equation}
u_{11}(r) =
4\varepsilon\left[\left(\frac{\sigma}{r}\right)^{12}-\left(\frac{\sigma}{r}\right)^6\right].
\label{lj}
\end{equation}
with a matrix-fluid interaction also given by a LJ potential, but with
an energy parameter $\varepsilon_{01} = \varepsilon/2$. 

For computational efficiency, we have truncated and shifted the
interactions at $R_c=10\sigma$ in both cases.

\section{Theory}

As mentioned, an explicit description of the
structure of a fluid inclusion in a particular matrix configuration can be achieved by
means of an approximation to the full two dimensional solution on the
Ornstein-Zernike (OZ) equation following
the prescription of Beglov and Roux\cite{JCP_103_360} (see below). A key element
in this description is the approximation of the fluid-fluid direct
correlation function, for which in Ref.~\onlinecite{JCP_103_360} that
of the bulk fluid was used. Whereas this may well be a
good approximation for the study of solvation of 
molecules\cite{Kovalenko1998} it turns out to be inappropriate under
conditions of close confinement. In fact, in most of the cases we
have studied, the full two
dimensional OZ equation does not even converge when this approach is
used. Therefore, in the present instance,
the fluid-fluid correlation will be approximated by that of a confined
fluid in a disordered matrix, which is in turn averaged (via the ROZ-formalism) over
disorder.  
This latter problem is known since the early nineties\cite{JStatPhys_1988_51_537,JCP_1992_97_4573,PRE_1993_48_233} to be amenable to
a theoretical description in terms of the ROZ equations. Here, the
matrix is manufactured by means of a templating procedure, which can be also
theoretically modeled with  Zhang and van
Tassel's\cite{JCP_2000_112_3006} ROZ formulation, which will be summarized in \ref{tROZ}
below.

\subsection{The full two dimensional OZ approach}

Following the work of Beglov and Roux\cite{JCP_103_360} we can
actually express the inhomogeneous density of a fluid under the
influence of an external field created by a set of porous matrix
particles in terms of an HNC-like expression of
the form
\begin{equation}
\rho_1({\bf r}) = \bar{\rho}_1\exp \left[ -U_{01}({\bf r})/k_BT + \int c_{11}({\bf r}-{\bf
    r}')(\rho_1({\bf r}')-\bar{\rho}_1) d{\bf r}\right]
\label{HNC2D}
\end{equation}
where $\bar{\rho}_1$ is an effective fluid density, whose connection
with the average fluid density, $\rho_1 = N_1/A$, ($N_1$ being the
number of fluid particles and $A$ the sample
area) will be defined below.  Eq. (\ref{HNC2D}) recalls Percus' source particle
approach\cite{Percus1964}, where one would take a matrix particle as
source of an external potential $U_{01}(r)$ and hence $\rho_1({\bf r}) =
\bar{\rho}_1g_{01}({\bf r})$, and the convolution within the
exponential accounts for the matrix-fluid indirect correlation
function, i.e. matrix-fluid correlations mediated by fluid
particles. Note that here, however, the external potential stems from
all matrix particles, by which, for a given matrix configuration
$\{{\bf r}_0\}\equiv \{ (x_0, y_0)\}$ with $N_0$
matrix particles, we have
\begin{equation}
U_{01}(x,y) = \sum_{i=1}^{N_0} u_{01}(x-x_{0_i},y-y_{0_i})
\end{equation}
with $u_{01}(r)$ given by Eq.~(\ref{u01}). Beglov and Roux\cite{JCP_103_360}
  approximate the fluid-fluid inhomogeneous correlation function by that of the bulk fluid, an approach which even if it somehow works for the
crude zeolite model studied in Ref. \onlinecite{JCP_103_360} is not
suitable here, as mentioned before.

 As an alternative, we propose the
use of the fluid-fluid correlations
approximated by the ROZ-HNC, by which $c_{11}({\bf r}-{\bf r}')$ in
Eq.~(\ref{HNC2D}) is given
by
\begin{equation}
c_{11}(x-x',y-y') = c_{11}^{{\rm ROZ-HNC}}(((x-x')^2+(y-y')^2)^{1/2})
\label{cff}
\end{equation}
and $c_{11}^{{\rm ROZ-HNC}} (r)$ is computed by solution of Eqs. (\ref{roz}) and
(\ref{cr}) below. Once $c_{11}(x, y))$ is known from Eq. (\ref{cff}) is known, Eq. (\ref{HNC2D})
can be solved iteratively using a mixing iterates
approach\cite{JCP_103_360}. To that purpose this relation is conveniently
rewritten as
\begin{equation}
h(x,y) = \exp\left[-U_{01}(x,y)/k_BT + \bar{\rho}_1\int dx'dy'
  c_{11}(x-x',y-y')h(x',y')\right]-1.
\label{hxy}
\end{equation}
and
\begin{equation}
\rho_1(x,y) = \bar{\rho}_1(h(x,y)+1).
\label{rho1}
\end{equation}

Note that the convolution in (\ref{hxy}) can easily be evaluated in
Fourier space, and the computations of the numerical Fourier transform
is  straightforward using efficient library routines such as those of the FFTW3
library\cite{frigo2005}. For the particular nature of our problem, and
taken into account that we will analyze fluid density distributions
for a given configuration of matrix particles that is assumed to have
periodic boundary conditions, the Fourier transforms can be carried
out without zero padding. Note that our calculations will be compared to
simulation results obtained using precisely the same periodic
boundary conditions. This periodic nature of the problem must be
very specially born in mind when approximating the inhomogeneous
direct correlation function in Eq.~(\ref{cff}) using the
averaged ROZ fluid-fluid correlations.

 Additionally, Eq.~(\ref{hxy}) can be linearized to yield a
Percus-Yevick (PY) like approximation of the form
\begin{equation}
h(x,y) = \exp(-U_{01}(x,y)/k_BT)\left[1 + \bar{\rho}_1\int dx'dy'
  c_{11}(x-x',y-y')h(x',y')\right]-1.
\label{hxypy}
\end{equation}

Even if the main result of the equation (\ref{hxy}) (in combination with a sum rule specified below) is via Eq. (\ref{rho1}) the {\it fluid}
spatial density distribution, $\rho_1(x,y)$, this quantity can also give
information on the average {\it matrix-fluid} correlations. In fact, these can be
obtained by means of 
\begin{equation}
g_{01}(r) = \frac{1}{\bar{\rho}_1N_0}\sum_{i=1}^{N_0} \int \rho_1({\bf
    r}-{\bf r}_{0_i}) d\theta_r
\label{g01}
\end{equation}
where $\theta_r$ is the polar angle of ${\bf r}$ and the summation
runs over all matrix particles.

Now, in Eq.~(\ref{hxy}), the value of the effective density $\bar{\rho}_1$
is not straightforwardly defined for our problem. In 
solvation problems\cite{JCP_103_360,Beglov1997,Kovalenko1998,Yamazaki2011} the
density in question can accurately be approximated by the bulk
density, but this is certainly not the case in situations of strong
confinement. On the other hand, we know that the density distribution must
satisfy the sum rule
\begin{equation}
\rho_1 = \frac{\bar{\rho}_1}{L_xL_y}\int dxdy(h(x,y)+1)
\label{rhobar}
\end{equation}
where $L_x, L_y$ are the dimensions of the periodic cell. Once the the
fluid density $\rho_1(x, y)$ is known, the effective homogeneous density $\bar{\rho}_1$
can be evaluated iteratively solving Equations (\ref{hxy}) and
(\ref{rhobar}) self-consistently.

\subsection{ROZ equations in a templated matrix}
\label{tROZ}
Following Zhang and van
Tassel\cite{JCP_2000_112_3006}, the ROZ equations for a templated
matrix can be derived from those of  a two-component
matrix  system simply dismissing the
interactions between the template
  and the adsorbed fluid after quenching. The explicit procedure to
build and solve the ROZ equations for our system of interest can be
found in Ref.~\onlinecite{JCP_2010_132_104705}, and we briefly sketch here
its main points for completion. 

The  ROZ equations can be  written in matrix form
in Fourier space in terms of density scaled Fourier transformed functions as\cite{JCP_2010_132_104705}
\begin{eqnarray}
\bm{H}^{01} &=& \bm{C}^{01} +
  \bm{C}^{00}\bm{H}^{01}+\bm{C}^{01}\bm{H}^{11}-\bm{C}^{01}\bm{H}^{12}\nonumber\\
\bm{H}^{11} &=& \bm{C}^{11} +
  \bm{C}^{10}\bm{H}^{01}+\bm{C}^{11}\bm{H}^{11}-\bm{C}^{12}\bm{H}^{12}\nonumber\\
\bm{H}^{12} &=& \bm{C}^{12} +
  \bm{C}^{10}\bm{H}^{01}+\bm{C}^{11}\bm{H}^{12}+\bm{C}^{12}\bm{H}^{11}-2\bm{C}^{12}\bm{H}^{12}\label{roz}
\end{eqnarray}
together with the decoupled matrix equation
\begin{equation}
\bm{H}^{00} = \bm{C}^{00}+\bm{C}^{00}\bm{H}^{00}
\label{ozm}
\end{equation}
where the superscript 0 and 1 denote the matrix and the fluid
respectively, and 2 the replicas of fluid particles. Now, each
of the matrix functions $\bm{F}^{ij}$ (where $\bm{F}$ stands for
either $\bm{H}$ or $\bm{C}$) can be explicitly expressed in terms of the
density scaled Fourier transforms of the total correlation function,
$\tilde{h}_{\alpha\nu}$, or the direct correlation function,
$\tilde{c}_{\alpha\nu}$, according to
\begin{equation}
\bm{F}^{01} = \left(\begin{matrix}\tilde{f}_{0_{\alpha}1}\\\tilde{f}_{0_{\beta}1}
  \end{matrix}
  \right)\;, 
\bm{F}^{11} = \tilde{f}_{11}\;,
\bm{F}^{12} = \tilde{f}^r_{11},
\end{equation}
and correspondingly for the matrix
\begin{equation}
\bm{F}^{00} = \left(\begin{matrix}\tilde{f}_{0_{\alpha}0_{\alpha}}&\tilde{f}_{0_{\alpha}0_{\beta}}\\\tilde{f}_{0_{\beta}0_{\alpha}}&\tilde{f}_{0_{\beta}0_{\beta}}
  \end{matrix}
  \right).\
\end{equation}
In the equations above, the superscript $r$ specifies
correlations between the replicas
of the  annealed fluid. Additionally we have $\bm{F}^{10}={\bm{F}^{01}}^T$,
where the superscript $T$ denotes the matrix transpose.

These equations in Fourier space are
complemented by the corresponding closures in $r$-space, which in the HNC
approximation read
\begin{eqnarray}
h_{11}(r) &=& \exp\left(-\beta
  u_{11}(r)+h_{11}(r)-c_{11}(r)\right)-1\nonumber\\
h_{{0_{\alpha}} 1}(r) &=& \exp\left(-\beta
  u_{{0_{\alpha}} 1}(r)+h_{{0_{\alpha}} 1}(r)-c_{{0_{\alpha}} 1}(r)\right)-1\nonumber\\
h_{{0_{\beta}} 1}(r) &=& \exp\left(h_{{0_{\beta}} 1}(r)-c_{{0_{\beta}} 1}(r)\right)-1\nonumber\\
h^r_{11}(r) &=& \exp\left( h^r_{11}(r)-c^r_{11}(r)\right)-1\label{cr}
\end{eqnarray}
where $f_{0_i1}=f_{1 0_i}$. For the matrix, we also have
\begin{equation}
h_{0_i0_j}(r) = \exp\left(-\beta
  u_{0_i0_j}(r)+h_{0_i0_j}(r)-c_{0_i0_j}(r)\right)-1\label{crmat}
\end{equation}
where the interaction between the matrix components, before the
template is removed, is given by Eq.~(\ref{vmm}). Eqs.~(\ref{crmat}) and
(\ref{ozm}) can be solved independently. As to Eqs.~(\ref{roz}), for
computational convenience they can be cast into a more compact matrix
form
\begin{equation}
\left(\begin{matrix}\bm{C}^{01}\\ \bm{C}^{11}\\ \bm{C}^{12}
  \end{matrix}
  \right) = \left(\begin{matrix} \bm{I}-\bm{C}^{00} & - \bm{C}^{01} &
      \bm{C}^{01} \\
 - \bm{C}^{10} &  \bm{I}-\bm{C}^{11} & \bm{C}^{12} \\
 - \bm{C}^{10} & -\bm{C}^{12} & \bm{I}-\bm{C}^{11}+2\bm{C}^{12} \end{matrix}
  \right)\left(\begin{matrix}\bm{H}^{01}\\ \bm{H}^{11}\\ \bm{H}^{12}
  \end{matrix}\right)
\label{roz2}
\end{equation}
where $\bm{I}$ is the identity matrix. Eq.~(\ref{roz2}) can be
efficiently inverted
for the components of the total correlation function in terms of the
direct correlation function using a LU-decomposition based algorithm\cite{laug}. 
Eqs.~(\ref{cr}) and (\ref{roz2}) can now be solved iteratively.

Once the correlation functions are determined, we can calculate
thermodynamic properties for the adsorbed fluid. A first quantity
that can be evaluated is the excess internal energy per 
particle (including both adsorbate and matrix particles),
\begin{equation}
\beta U_1/N =
\frac{1}{2}\frac{\rho_1\rho_1}{\rho}2\pi\int
dr r g_{11}(r)\beta u_{11}(r) +
\frac{\rho_1\rho_{0}}{\rho}2\pi\int dr r g_{01}(r)\beta u_{01}(r)
\label{uex}
\end{equation}
with $\rho = \rho_0+\rho_1$. Finally, the ROZ-HNC direct expression
for the chemical potential
is\cite{JPCB_2001_105_4727,JCP_1992_97_8606,JCP_1999_111_10275} 
\begin{eqnarray}
\beta\mu_1 &=& -\sum_{i=\alpha,\beta}\rho_{0_i}\tilde{c}_{0_i1} (0) -\rho_{1}(\tilde{c}_{11} (0)-\tilde{c}_{11}^r(0))+
\frac{1}{2}\sum_{i=\alpha,\beta}\rho_{0_i}2\pi\int d r r
h_{0_i1}(r)\gamma_{0_i1}(r)\nonumber\\
& & + \frac{1}{2}\rho_12\pi\int dr r(h_{11} (r)\gamma_{11} (r)-h_{11}^r (r)\gamma_{11}^r (r))+\log (\rho_1\Lambda_1^2)\label{mu}
\end{eqnarray}
where $\Lambda_1$ is the de Broglie wavelength for the fluid
particles, and  $\gamma(r) = h(r) - c(r)$. 

\section{Results}
As a first test of our approach, we have checked the performance of
the ROZ-HNC equation for the matrix density
$\rho_0\sigma^2=0.314$. This corresponds to the highest density for
which the HNC equation can be solved for the non-additive hard disk
fluid which is the precursor of our templated matrix. The HNC
matrix-matrix correlations enter the solution of the ROZ-HNC equations
through Eq. (\ref{roz2}), and therefore, in all theoretical
calculations in this work, we will approximate the confined
fluid-fluid correlations by those of the ROZ-HNC solved for a matrix
of $\rho_0\sigma^2=0.314$, even when the case of study has a somewhat
larger matrix density and a different topology as will be illustrated
below. 

As mentioned, matrix configurations were generated from a symmetric
mixture of non-additive hard spheres. For each matrix configuration,
the SALR fluid is inserted in 
the matrix using a Grand Canonical Monte Carlo simulation (GCMC),
generating half a  million fluid configurations (each configuration corresponds
to one particle insertion/deletion attempt, and $N_1$ displacement trials,
where $N_1$, as before, is the number of fluid particles for the configuration in question),
and then the results are averaged over ten matrix configurations. The
ROZ equations have been solved following the procedure introduced in
Ref.~\onlinecite{JCP_2010_132_104705} and with the same discretization
conditions. 

\begin{figure}
  \includegraphics[width=13cm,clip]{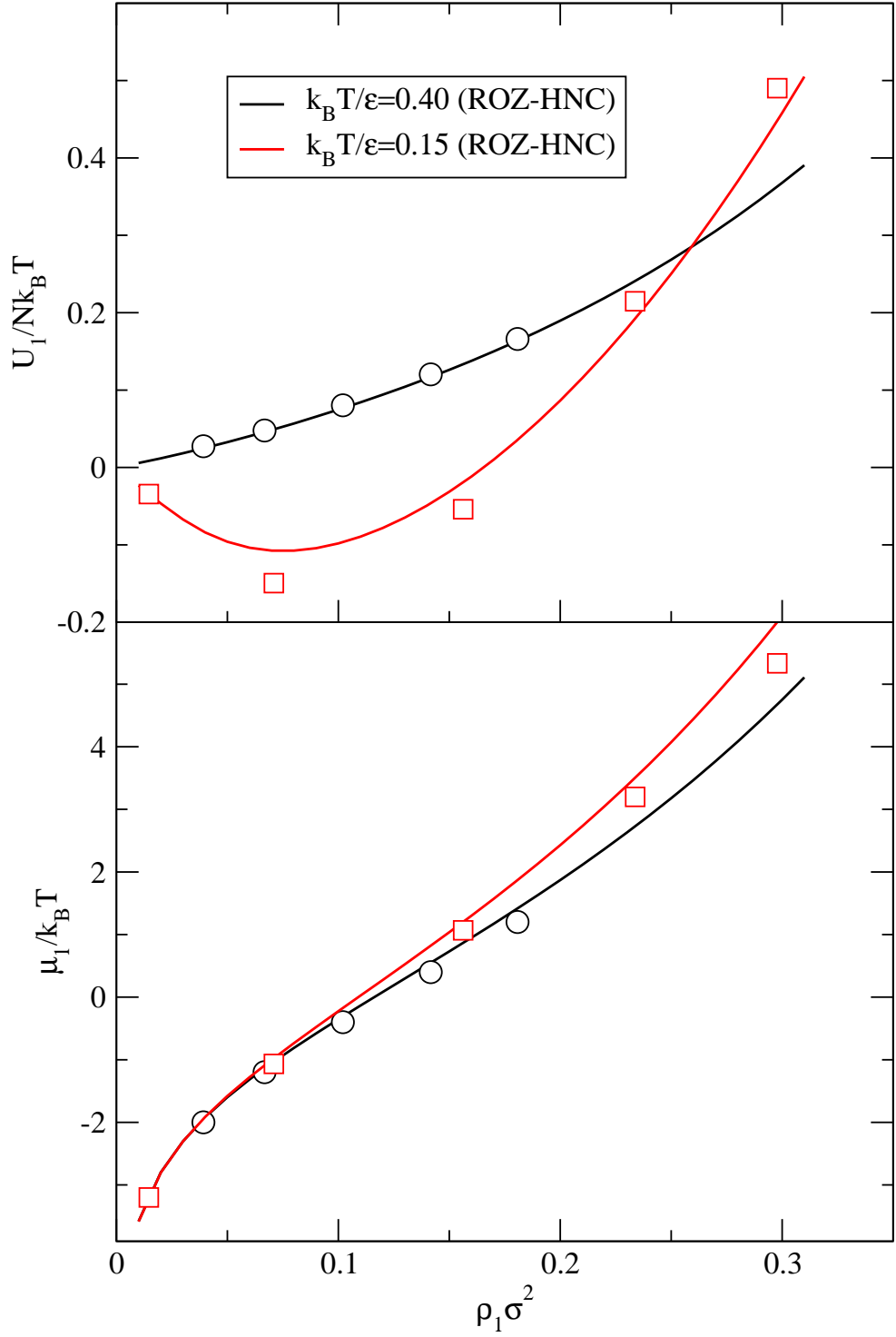}
\caption{Average fluid excess internal energy and chemical potential
  for a matrix density $\rho_0\sigma^2=0.314$ at various
  temperatures. \label{thermoaver}}
\end{figure}

\begin{figure}
  \includegraphics[width=13cm,clip]{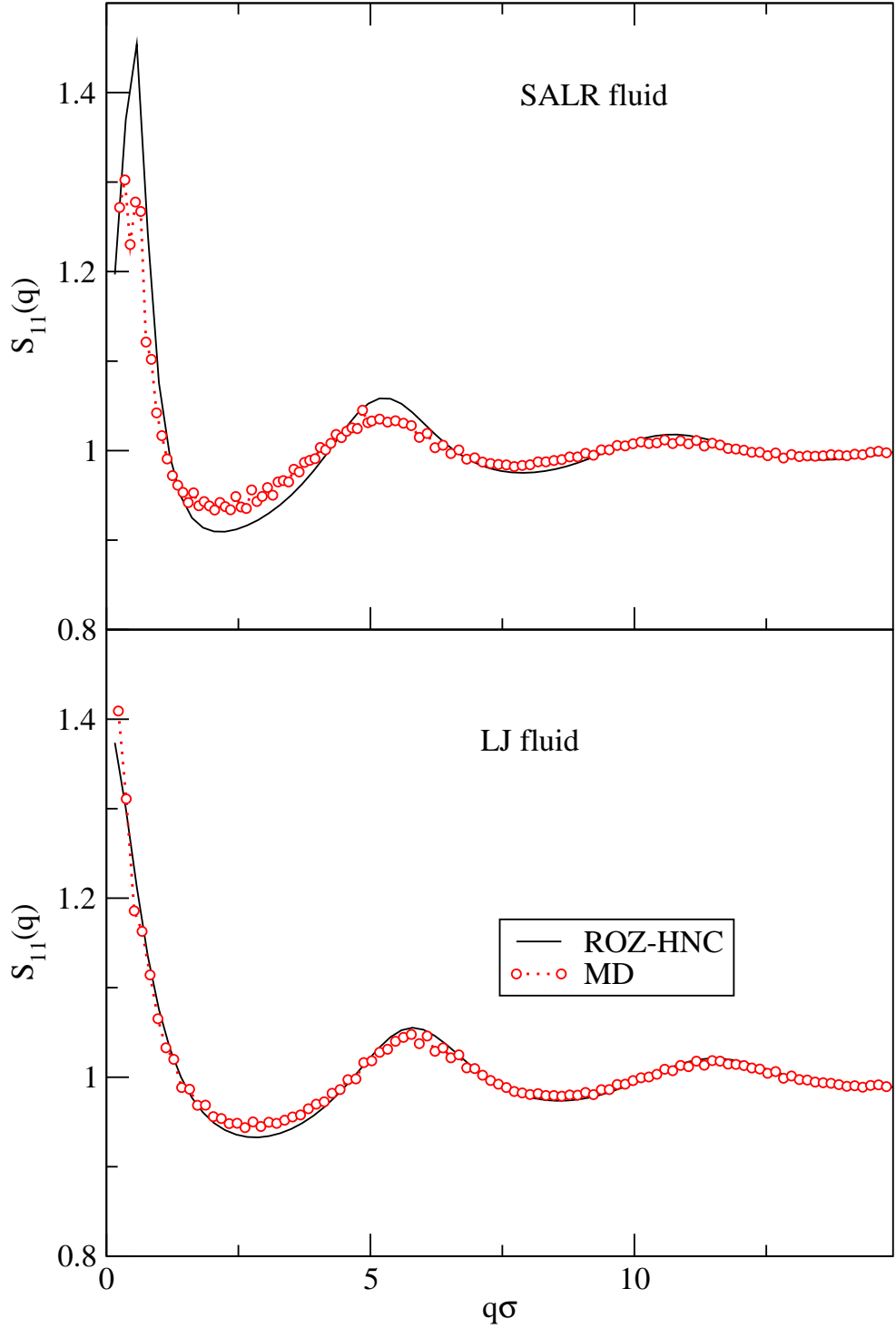}
\caption{Fluid-fluid structure factor as calculated from the ROZ-HNC
  equation (curves) and from MD (symbols) for the adsorbed SALR fluid (upper graph) and
  the LJ system (lower graph) for $\rho_1\sigma^2=0.0596$ and
  $\rho_0\sigma^2=0.324$.  Note the marked pre-peak at $q=0.578\sigma$ 
 for the SALR fluid. The
  LJ fluid $S_{11}(q)$ grows as $q\rightarrow 0$, which indicates the
  vicinity of the condensation transition.\label{sQ}}
\end{figure}

 In
Figure \ref{thermoaver} we plot the adsorption isotherms (lower curve)
and the excess potential energy for a relatively high temperature
($k_BT/\varepsilon=0.4$) and a much lower one
($k_BT/\varepsilon=0.15$), for which clustering effects can be
appreciated. Actually, the low density non-monotonous behavior of the
internal energy reflects the competition between attractive forces
(dominant at low densities) and repulsive forces that start to
shape the system's behavior for densities above $\rho_1\sigma^2 \approx0.06$. 
A signature of clustering can be appreciated in Figure \ref{sQ}
where the fluid-fluid structure factor, $S_{11}(q)$  calculated from the ROZ-HNC
equation is compared with the one extracted from the simulation.  This
quantity is defined as
\begin{equation}
S_{11}(q) = 1+\rho_1\tilde{h}_{11}(0)=\frac{1}{N_1} \left<|\sum_{i=1}^{N_1}
e^{i{\bf q}{\bf r}_i}|^2\right>
\label{sq}
\end{equation}
where $N_1$ is the number of fluid particles and $<\ldots >$ denotes
the ensemble average. In Eq.~(\ref{sq}), the fluid-fluid correlation function must be replaced (in the ROZ formalism) by its
connected counterpart, $h_{11}(r)-h_{11}^r(r)$, when the average over
disorder is performed. 

The simulation results presented in the Figure \ref{sQ} correspond to
a molecular dynamics (MD) run
for a single matrix configuration and in which the initial fluid
configuration is generated in a GCMC run. MD results correspond to
averages carried out for one to two million configurations, using
samples with 441 and 2399 fluid and matrix particles respectively and
an integration time step of 0.0025 in reduced time units. The matrix
density in this case is $\rho_0\sigma^2=0.324$, slightly above the one
used in the solution of the ROZ-HNC equations. 
The presence of intermediate
range order\cite{Godfrin2013}, which in our case can be identified
with clustering, is indicated by the marked pre-peak in the fluid-fluid
structure factor at 
$q=0.578\sigma^{-1}$, that reflects intercluster correlations for
distances around $11\sigma$. For comparison in the lower graph we
present results obtained using the same matrix and initial fluid
configuration but  with interactions of a LJ fluid, Eq. (\ref{lj}). We observe in the latter case that the structure
factor lacks any signature of intermediate range order, but it clearly
shows signs of an approaching divergence at $q\rightarrow 0$, i.e. the
vicinity of the condensation transition. 

\begin{figure}
\includegraphics[width=13cm,clip]{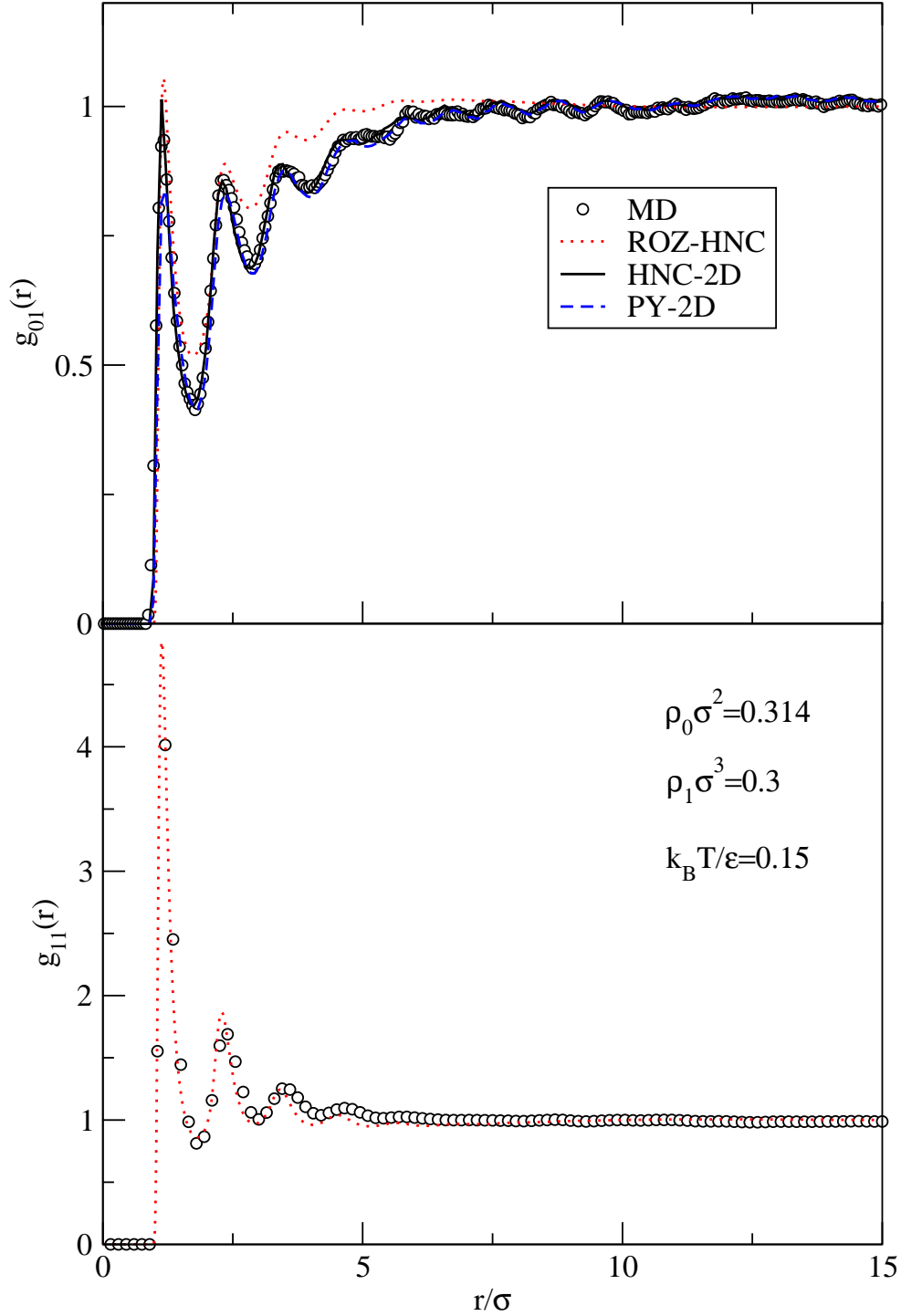}
\caption{Average matrix-fluid and fluid-fluid correlations for a SALR
  fluid inclusion as
  estimated from MD simulations run from a given GCMC configuration
  with a fixed matrix configuration 
  and by means of of PY-2D and HNC-2D equations and ROZ-HNC
  equations. Note that the ROZ equations provide the average of the
  correlations over matrix disorder. \label{gct015}}
\end{figure}

\begin{figure}
\includegraphics[width=13cm,clip]{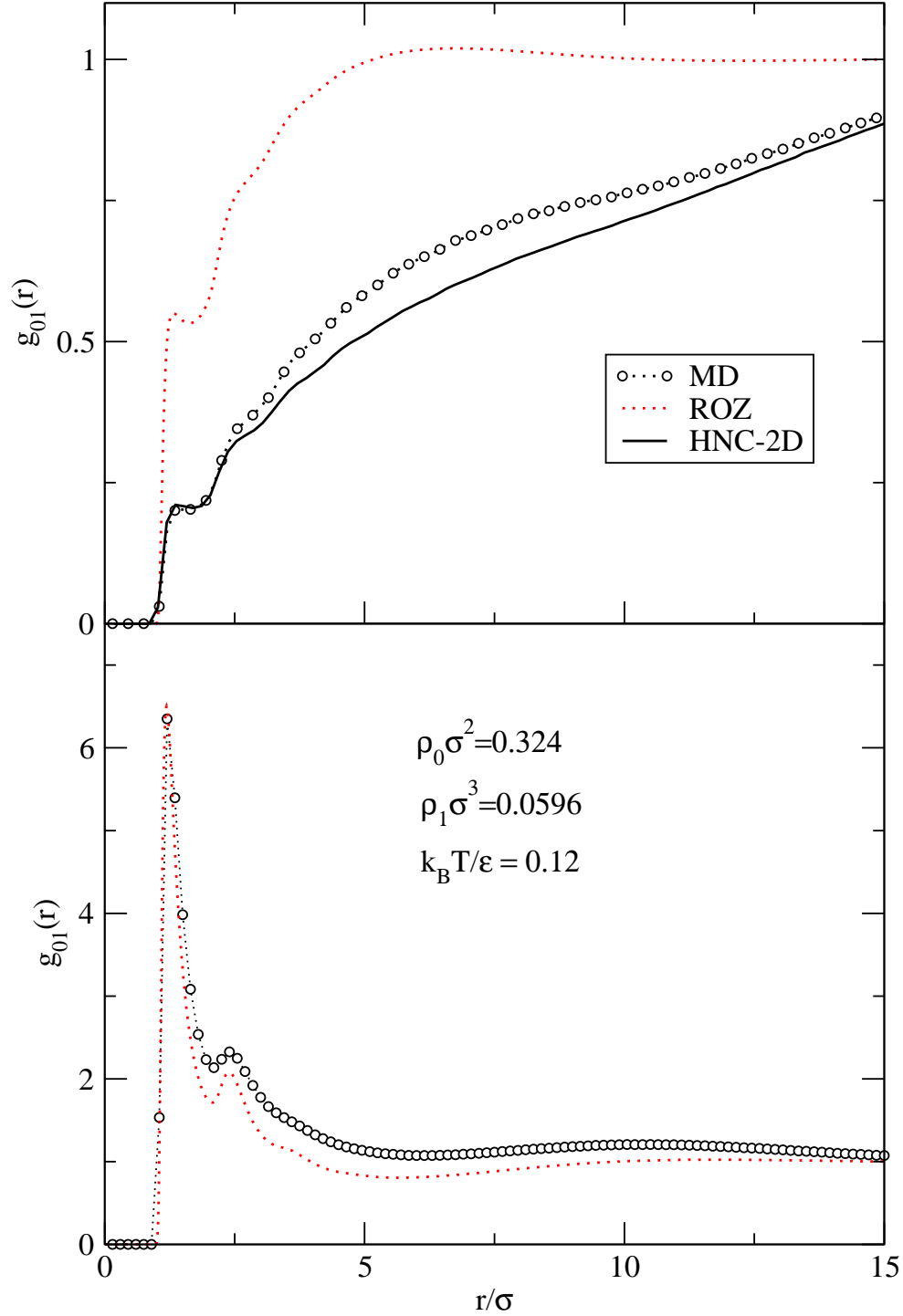}
\caption{Same as figure \ref{gct015} for a much lower fluid density
  for which clustering effects are more apparent, as can be inferred
  by the long range of the fluid-fluid correlations,
  $g_{11}(r)$, which exhibit a wide maximum around $11\sigma$.\label{gct012}}
\end{figure}

In the lower graphs of
Figures \ref{gct015} and \ref{gct012}, we can see the performance of
the ROZ-HNC equation for the calculation of the fluid-fluid
correlations, which is relatively good for the high density case
($\rho_1 \sigma = 0.3$, cf. Fig.~\ref{gct015}) and acceptable for the low density
($\rho_1 \sigma = 0.0596$, cf. Fig.~\ref{gct012}). Differences can be in part attributed to the
fact that we are comparing the ROZ results obtained for a matrix density
$\rho_0\sigma^2=0.314$ with those of the simulation sample for which
$\rho_0\sigma^2=0.324$, with the additional modification induced in
the matrix topology by the removal of disconnected matrix particles. Notice that
both theory and simulation reproduce the presence of a wide maximum at
approximately $11\sigma$, in correspondence with the location of the pre-peak in
$S_{11}(q)$. The comparison of the ROZ-HNC data with simulation results considerably
worsens for the fluid-matrix 
correlations (upper graphs in the same figures), particularly at low
density where clustering is more evident. Obviously, this discrepancy
results from the poor description of matrix-matrix correlations when
using $\rho_0\sigma^2=0.314$ ROZ results to model those of
$\rho_0\sigma^2=0.324$, which is particularly crucial for the $g_{01}(r)$
correlations at low fluid densities. For higher fluid densities, the
packing effects of the fluid dominate and this explains why the ROZ
performance for $\rho_1\sigma^2=0.3$ is far better than
for $\rho_1\sigma^2=0.0596$, as far as matrix-fluid correlations are
concerned.

Nonetheless,  we will only need the fluid-fluid correlations $g_{11}(r)$ to solve
the HNC-2D equation (\ref{hxy}), and those are reasonably
approximated by the ROZ-HNC. The solution of the HNC-2D  is done using a discretized
two dimensional grid of $N_x\times N_y$ points (in this case
$N_x=N_y=512$) with a grid spacing that is given by $\delta x =
L_x/N_x$ and $\delta y = L_y/N_y$, where $L_x$ and $L_y$ represent the
size of the simulation box corresponding to the matrix configuration whose
fluid density distribution will be calculated using
Eq.~(\ref{hxy}). Here, we will consider 
$L_x=L_y=86.066\sigma$ for low fluid density calculations
, $\rho_1\sigma^2=0.0596$, (with $\rho_0\sigma^2=0.324$) and $L_x=L_y=39.84\sigma$, for the
moderately high fluid density, $\rho_1\sigma^2=0.3$ (with $\rho_0\sigma^2=0.314$). The first
results that come out from the solution of the HNC-2D equation are the
matrix-fluid correlations that are depicted in the upper graphs of
Figures  \ref{gct015} and \ref{gct012}. It is evident that the full 2D
approach considerably improves upon the ROZ-HNC approximation for the
average $g_{01}(r)$ distributions for a specific matrix configuration,
 in particular at low densities. For the highest density we include in the Figures
results from the PY-2D approximation (\ref{hxypy}), which in this
particular instance is more or less of the same quality as the HNC.

\begin{figure}
\subfigure[MD]{\includegraphics[width=11cm,clip]{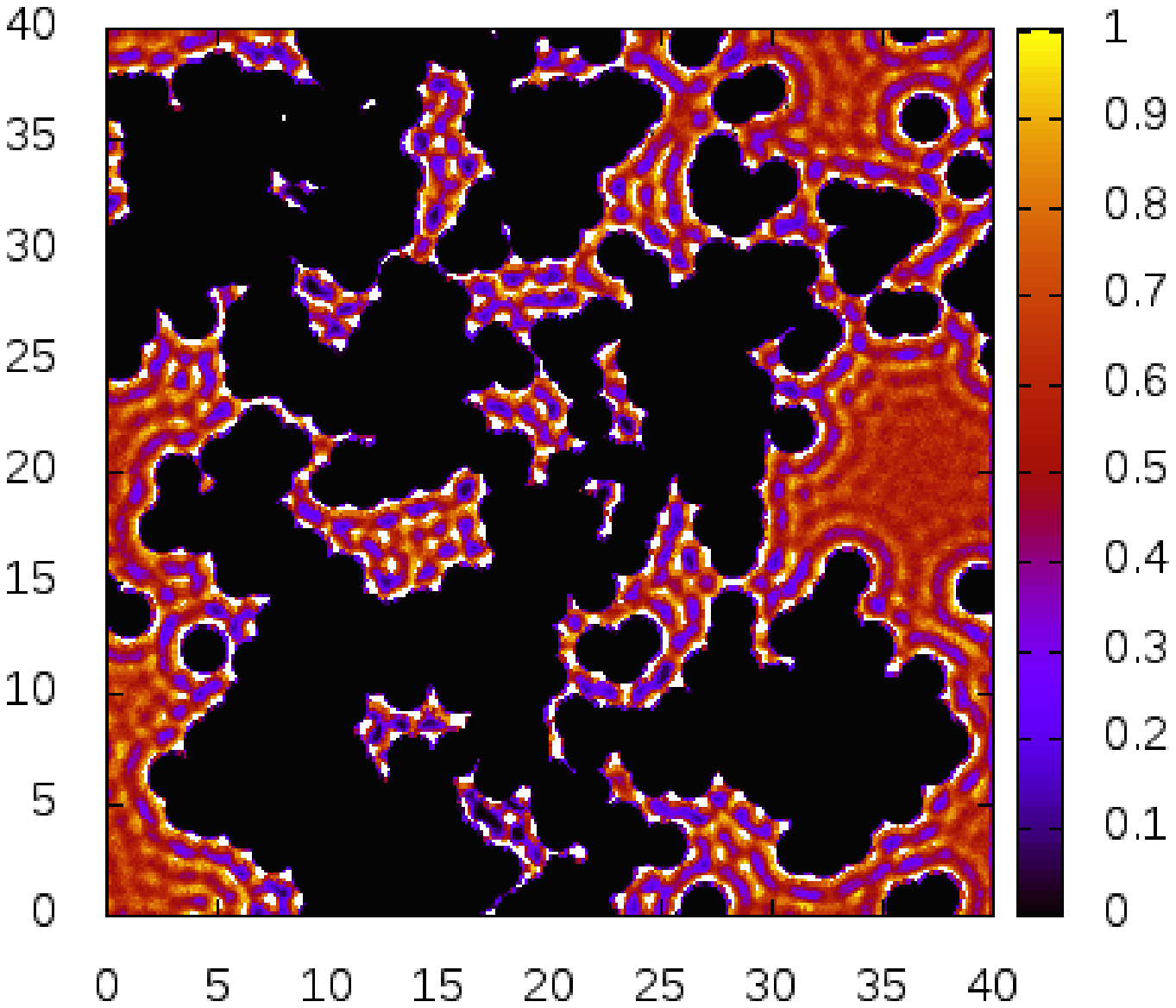}\label{MD015}}
    \subfigure[HNC]{\includegraphics[width=11cm,clip]{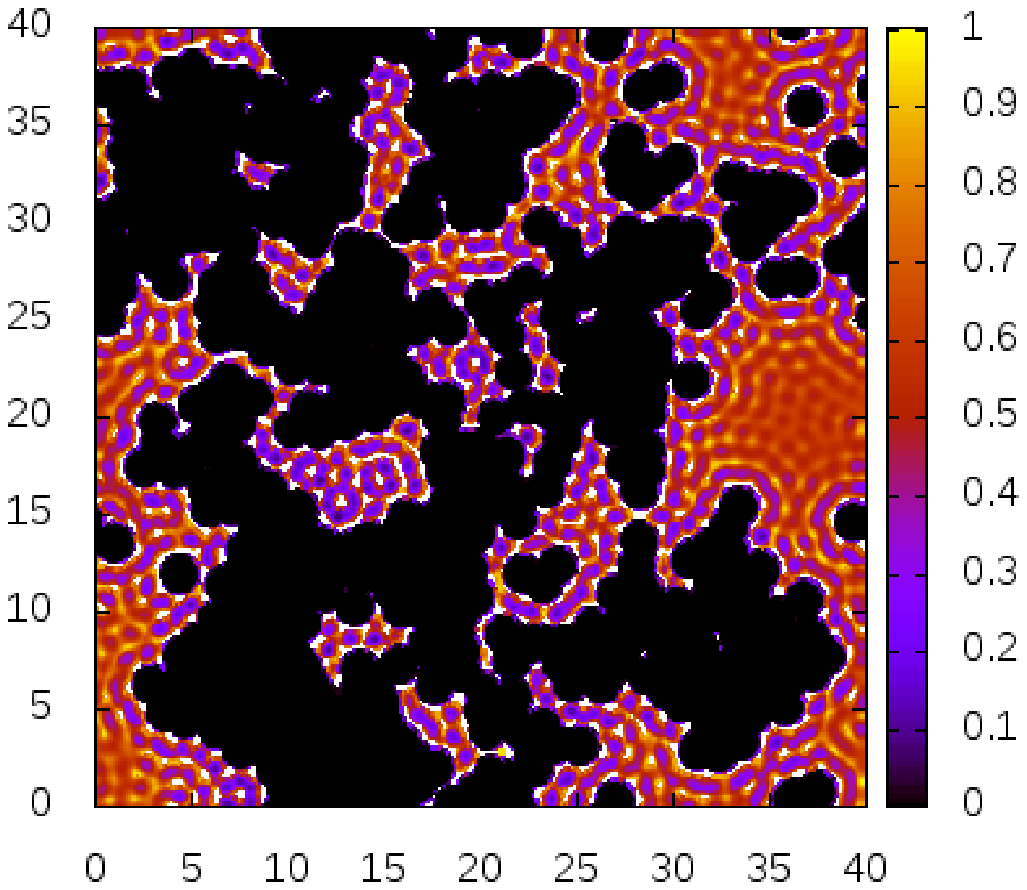}\label{HNC015}}
\caption{Fluid density distribution $\rho_1(x,y)$ for the SALRC fluid for
  $\rho_1\sigma^2=0.3$ and $k_BT/\epsilon=0.15$
  ($\rho_0\sigma^2=0.314$) from MD and in the
  HNC-2D approximation.\label{rhoxy03}} 

\end{figure}

\begin{figure}
\subfigure[MD]{\includegraphics[width=11cm,clip]{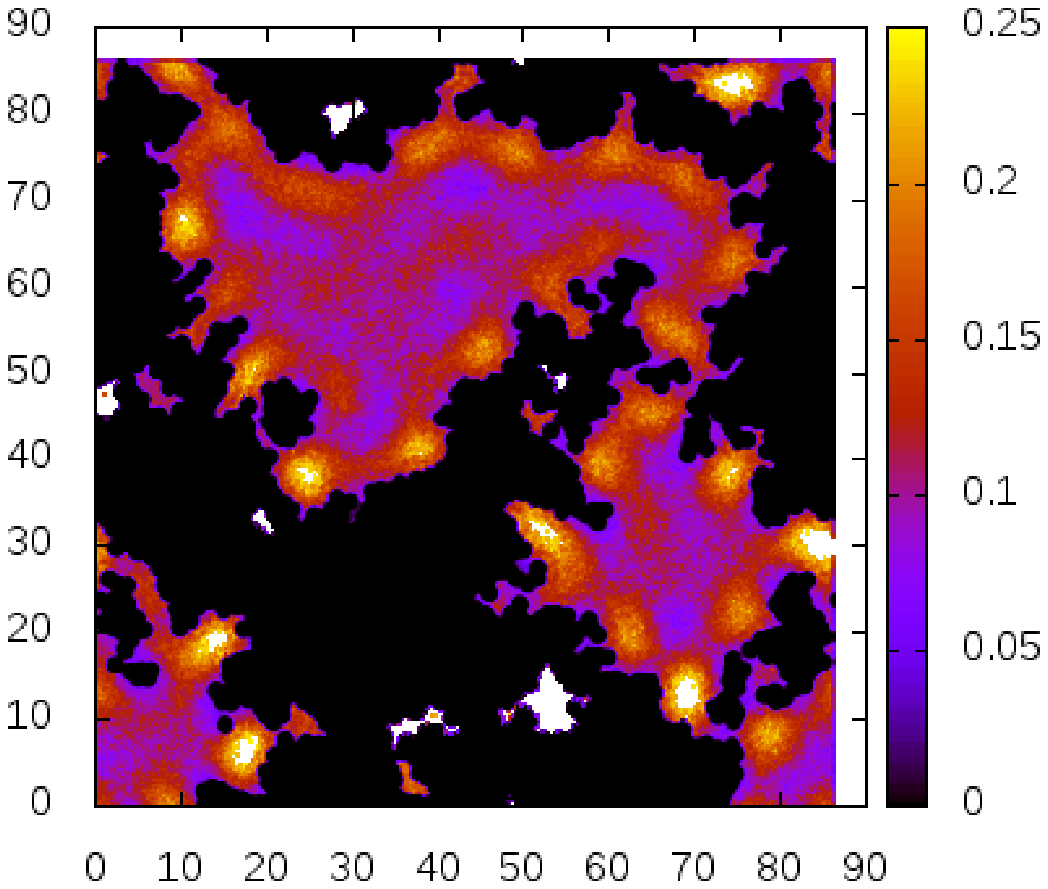}\label{MD012}}
    \subfigure[HNC]{\includegraphics[width=11cm,clip]{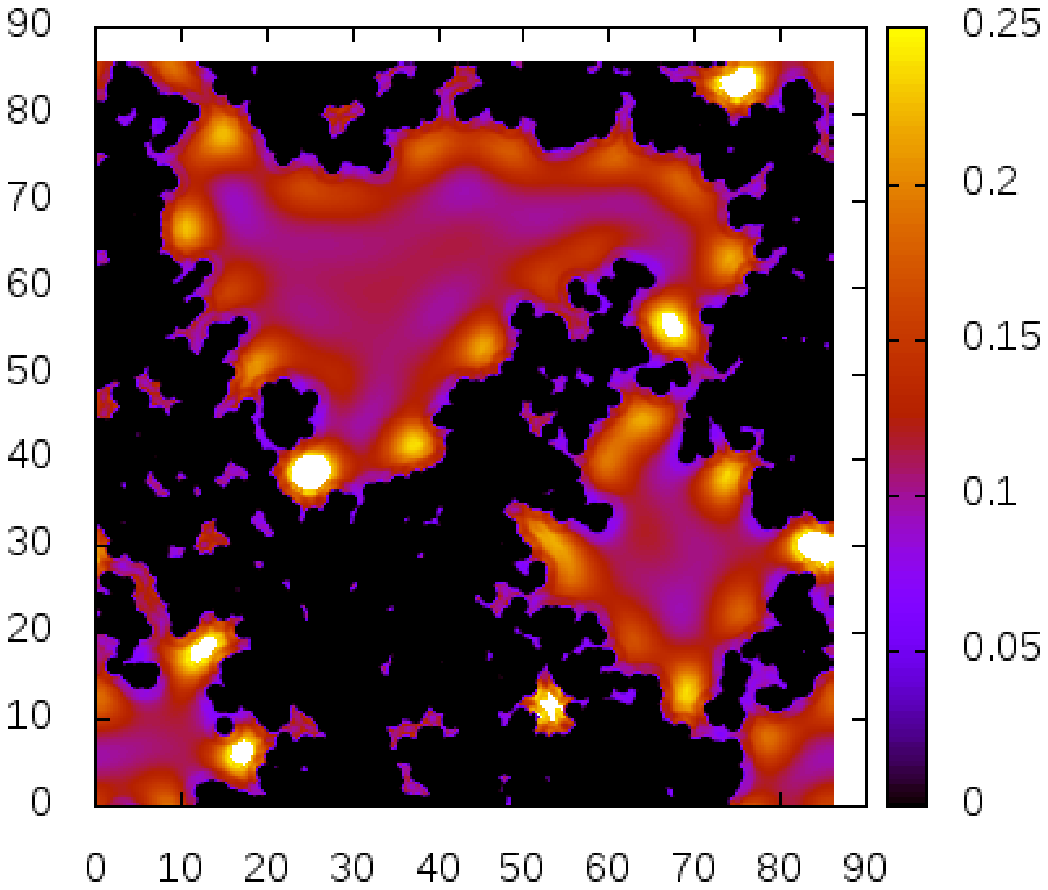}\label{HN012}}
\caption{Fluid density distribution $\rho_1(x,y)$ of the SALR fluid for
  $\rho_1\sigma^2=0.0596$ and $k_BT/\epsilon=0.12$
  ($\rho_0\sigma^2=0.324$) from MD and in the
  HNC-2D approximation. Regions for which $\rho_1(x,y) > 0.25$ are
  plotted in white.} 
\label{rhoxy006}

\end{figure}

Now in Figures \ref{rhoxy03} and \ref{rhoxy006} we present the explicit 2D
fluid density distributions $\rho_1(x,y)$, for $\rho_1\sigma^2=0.3$
($\rho_0\sigma^2=0.314$), and $\rho_1\sigma^2=0.0596$
($\rho_0\sigma^2=0.324$) for SALR fluids at temperatures where clustering becomes
apparent (particularly at low density). One can immediately appreciate what we have
commented upon above concerning the different matrix
topologies. Despite the relatively small difference in the density $\rho_0$, one can
see that the matrix porosity is appreciably larger for $\rho_0 \sigma^2 = 0.324$ (Figure
\ref{rhoxy006}), and therefore matrix-matrix correlations (and
hence matrix-fluid correlations as well) are quite different in
both instances as we have seen. This effect is less appreciable in the
fluid-fluid correlations, which are mostly conditioned by the effective
density of the adsorbed fluid (similar in both cases). 

 In the case of the higher fluid
density, in Figure  \ref{rhoxy03} one readily observes that the HNC-2D
equation actually reproduces quite well the simulated density
distribution, and interestingly the maxima and
minima of $\rho_1(x,y)$ are seen to display the features of a partly ordered
system (approaching the local structure of a triangular lattice). Note that in this figure the
black region represents the area of the system in which the fluid
density vanishes. In this case this is precisely the area inaccessible to the
centers of adsorbate atoms, i.e., the exclusion surface of the matrix as
defined by the positions of its constituent atoms and their
corresponding individuals exclusion surfaces. This exclusion surface
per matrix particle is given for 
  for our matrix-fluid interaction by $\approx \xi\pi\sigma^2$,
 with a parameter $\xi$ that accounts for the potential softness set
 to $\xi =0.98$ for the SALR potential, and $\xi =$0.8 for the LJ
 matrix-fluid interaction. In the case of the MD picture, the black region also includes
 those isolated matrix cavities that were empty from fluid particles
 in the starting configuration of the MD run. This empty cavities will
 be 
 thus indistinguishable from the area of the sample excluded by the
 matrix in these spatial fluid density maps.

It can  be
appreciated that in the HNC-2D picture there is a substantially larger number of
small disconnected pores partly filled with fluid inclusions as
compared with the MD results. In this latter instance most of the disconnected pores
happen to be empty for our particular initial fluid configuration. As mentioned,
this results from  the fact that our simulation  corresponds to a MD run started
from a single GCMC configuration of the fluid particles. A better agreement in
this respect would be reached at a much higher computational 
cost calculating the fluid density $\rho_1(x,y)$ as the average from a
series of MD runs started from different GCMC configurations of the fluids
sampling the same chemical potential. In this case, the average population of
the disconnected pores would be determined by the equilibrium grand
canonical partition function and we would not have to rely on a
single initial fluid configuration which might well be away from the
equilibrium average. Notice that for large pores, the subsequent MD
sampling would essentially converge toward the GCMC result. An
alternative approach, would be the use of configurations in which
fluid particles are not allowed to populate disconnected cavities in the
matrix, and modify consequently the matrix-adsorbate interaction in the HNC-2D
approach to include an artificial hard core potential that forces
$\rho_1(x,y)\rightarrow 0$ inside these isolated cavities
(e.g. retaining the template matrix particles that occupy these
cavities in the original non-additive hard disk mixture). For
simplicity, we have retained the original control pore glass like
interaction and used our initial test GCMC results as input to
generate the MD trajectories.

On the other hand, the reason for using MD generated configurations and not
the output of the GCMC directly, is simply related to computational efficiency
in the calculation of the simulated $\rho_1(x,y)$. In order to
obtain a smooth  density distribution, one needs to perform an
intensive sampling of the configurational space with a large number of
averages of spatially contiguous particle configurations, which is much
simpler to attain using MD. 

In Figure \ref{rhoxy006} we observe a density distribution $\rho_1(x, y)$
characteristic of the presence of clustering. One sees immediately that there are
spatially separated regions of substantially higher fluid density,
which tend to concentrate in pockets and bays of the porous structure,
despite the fact that the matrix-fluid interaction is purely repulsive. This is due
to the fact that, once particles aggregate in clusters as a
consequence of the short
range attractive part of the potential, the long-range repulsion between the
clusters pushes them apart to places where at least some of them are
``sheltered'' by matrix particles. The intercluster separation lies
in the range $10\sigma \sim 11\sigma$, in agreement with the position of the $S_{11}(q)$
pre-peak and the wide maximum of $g_{11}(r)$. The theoretical results are
in excellent agreement with the simulated density
distribution. Moreover, we have seen that as the simulation proceeds
the results tend to approach to the theoretical prediction, since the
reorganization of the clusters is a relatively slow process, 
particularly when they get trapped in narrow pockets. This is one of
the reasons why the MD simulations have to be exceptionally long to
yield a reliable $\rho_1(x,y)$. As an illustration,  the fluid density
inside the cavity at $\approx (12\sigma,55\sigma)$ in Figure
\ref{MD012} was found to remain well above $\rho_1(x,y)\sigma^2 > 0.3$ for close to one
million MD steps, to slowly decrease and reach $\approx 0.17$ after
another two million steps, much
closer to the predicted HNC-2D estimate seen in Figure \ref{HN012}. 

\begin{figure}
\includegraphics[width=12cm,clip]{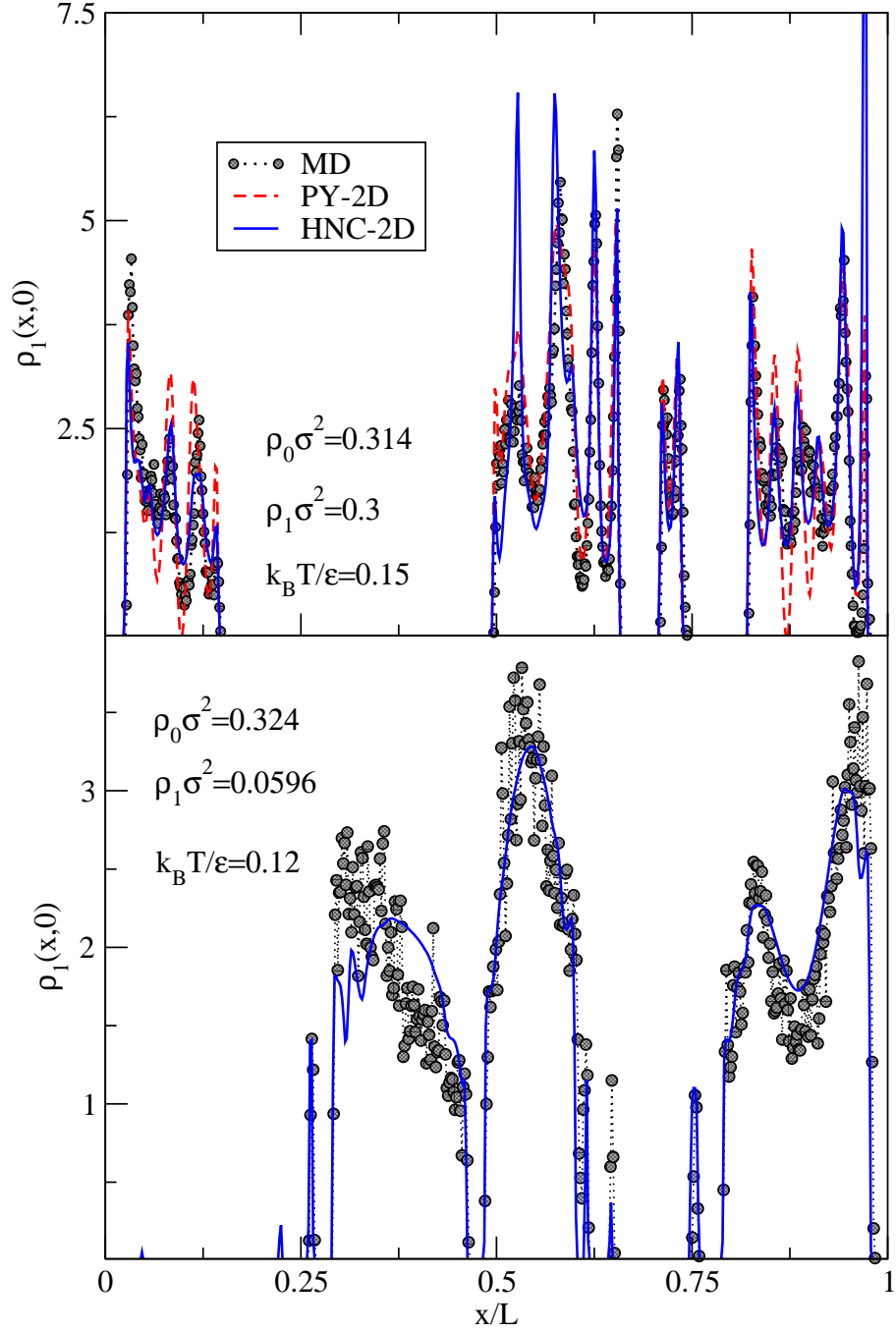}
\caption{Illustration of the density profile $\rho_1(x,0)$ of the SALR fluid along
  the $x$-axis taking as origin a given 
  matrix particle. The upper figure corresponds to the density map of
  Figure \ref{rhoxy03} and the lower one to the map of Figure
  \ref{rhoxy006}. Distances are scaled with the side of the simulation
box since each figure corresponds to samples of different size, and
the reference matrix particle is also a different one. \label{perfilxy}}
\end{figure}

In order to appreciate the performance of the integral equations more
quantitatively, in Figure \ref{perfilxy} we plot the density profiles
along the $x$-axis using as reference a given matrix particle. We
observe that the PY-2D approximation underestimates the value of the
density minima, and the HNC-2D slightly overestimates the maxima,
which is a characteristic feature of the HNC correlation
functions. Note that the theoretical profile exhibits some spikes
corresponding to fluid particle inclusions in the aforementioned
isolated cavities. These
spikes are either absent in the MD results or have a much lower
intensity, as a result of a much lower (or zero) initial density of fluid
particles in the cavity in the starting MD configuration. This is
again explained  in the preceding paragraph as a result of the use of a
single GCMC configuration as starting point for our MD
calculations. Aside from this detail, the agreement between the theory
and the simulation is remarkable. 

\begin{figure}
\includegraphics[width=14cm,clip]{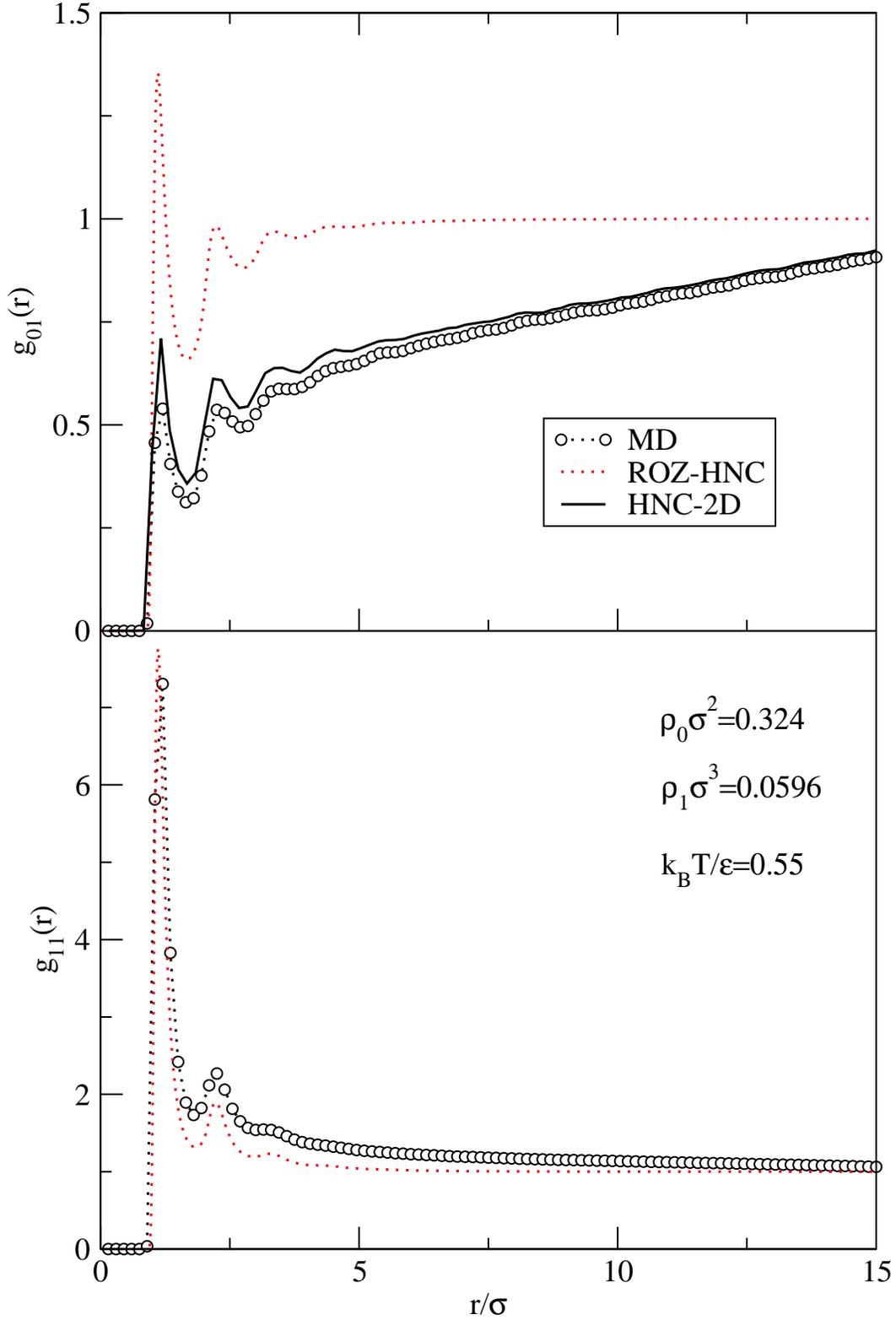}
\caption{Same as figure \ref{gct012} for a LJ fluid in a LJ
  matrix. Note the higher intensity of the first peaks as a consequence of
  the LJ attraction, the pronounced spatial structure of $g_{01}(r)$  and
  the lack of maximum at $10\sigma$ in $g_{11}(r)$ due to the absence of clustering.\label{gctLJ}}
\end{figure}

Finally, for the sake of comparison we have run a long MD simulation
(two million independent configurations in a run of ten million time steps) 
from the same starting fluid and matrix configuration as before ($\rho_0\sigma^2=0.324$), but
where now fluid-fluid and matrix-fluid interactions are truncated and
shifted LJ potential.  The temperature of the
run was set to $k_BT/\varepsilon=0.55$ and the fluid density as
before $\rho_1\sigma^2=0.0596$. These conditions are quite close to the
gas-liquid transition as can be inferred from the large $S_{11}(0)$
values in Figure \ref{sQ}. We have solved the ROZ-HNC
equations for this system and obtained the density distribution using
the HNC-2D equation. In Figure \ref{gctLJ} we show the fluid-matrix
and fluid-fluid correlations. As in the case of the low density SALR
fluid, again here the fluid-matrix correlation ROZ-HNC predictions are
rather poor, due to the inaccurate representation of the matrix-matrix
correlations for this matrix configuration. Also, fluid-fluid correlations do not show any trace of
intermediate long range ordering or clustering (the high values of the
first peak of $g_{11}(r)$ are just an indication of confinement and the
fluid-fluid correlation dies out rapidly). This is in clear contrast
with the long range features found in Figure \ref{gct012}. Again, the
HNC-2D fluid-matrix correlation agrees quite well with the MD
results. 

\begin{figure}
\subfigure[MD]{\includegraphics[width=11cm,clip]{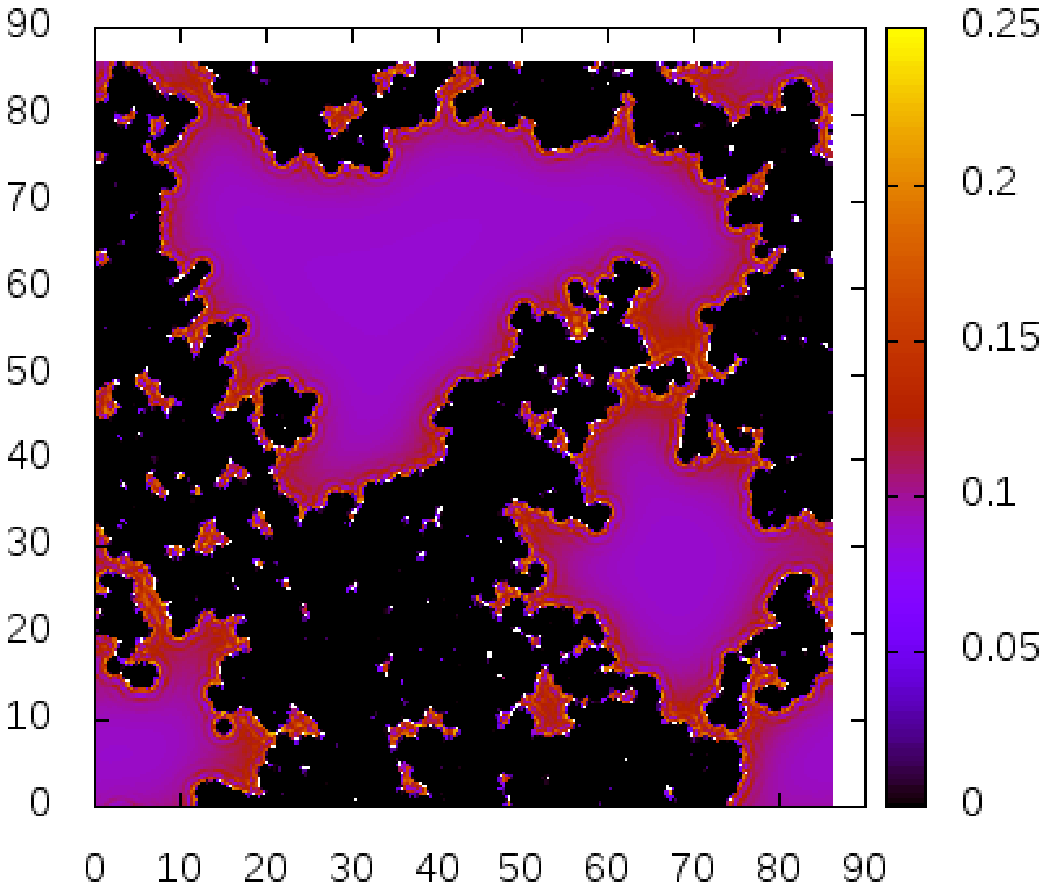}\label{MDLJ}}
    \subfigure[HNC]{\includegraphics[width=11cm,clip]{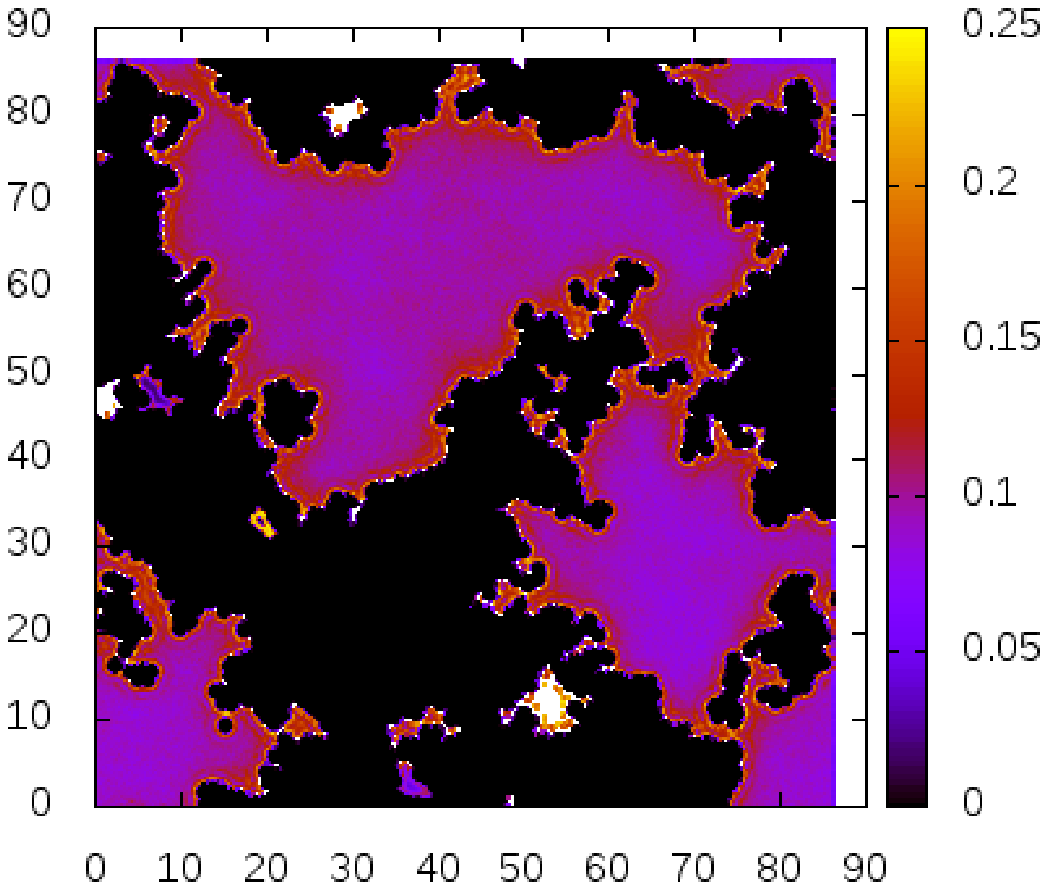}\label{HNLJ}}
\caption{Fluid density distribution $\rho_1(x,y)$ for
  $\rho_1\sigma^2=0.0596$ and $k_BT/\epsilon=0.55$
  ($\rho_0\sigma^2=0.324$) from MD and in the
  HNC-2D approximation for a LJ fluid inclusion in a matrix formed
  by LJ particles.} 
\label{rhoLJ}

\end{figure}

\begin{figure}
\includegraphics[width=14cm,clip]{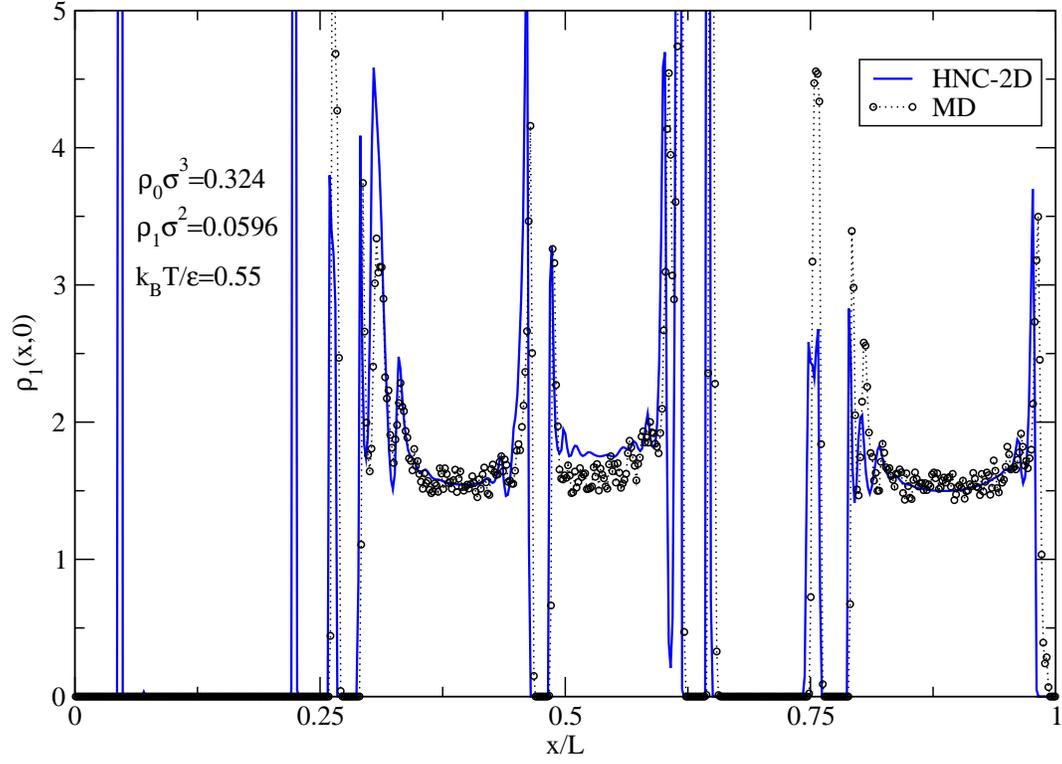}
\caption{Illustration of the LJ fluid density profile $\rho_1(x, 0)$ along the $x$-axis taking as origin a given
  matrix particle. The fluid-matrix attraction is reflected in the
  large values of the density profile in the immediate vicinity of the
  matrix particles.\label{perfilLJ}}
\end{figure}

Now, the fluid density distribution  for the same density as that of Figure
\ref{rhoxy006} can be seen for the LJ system in Figure \ref{rhoLJ}, and again we
appreciate a remarkable agreement between theory and simulation. In
this case, the simulation had to be particularly long for the 
density distribution to be smooth enough. The only salient feature
appreciated in Figure \ref{rhoLJ} is the fact that the fluid density
is enhanced near the pore walls, due to the attractive nature of the
matrix particles. During the simulation run one can see the formation
of short lived aggregates as a consequence of the vicinity of the liquid-gas
transition, but  in contrast to the SALR fluid, they have
no preferred positions within the pore space, and the local density enhancements average
out. Again, in Figure \ref{perfilLJ} we can have a more quantitative
appreciation of the quality of the results in the density profile
along the $x$-axis. The attractive nature of the pore particles is
evidenced by the large values of $\rho_1(x,0)$ near the matrix
particle boundaries, in contrast to the situation in Figure
\ref{perfilxy}, where  the repulsive nature of matrix-particle interaction used
for the SALR fluid simulations is evident.

In summary, we have explored the ability of a full 2D solution of the
HNC (and PY) equation for a fluid inclusion in a disordered porous
matrix with a large degree of porosity. The equation was complemented
by the use of a ROZ-HNC equation to faithfully  approximate the fluid-fluid
correlations under confinement. We have shown that this approach
reproduces in a satisfactory way the average fluid-fluid spatial correlations of different
types of adsorbates and its combination with the HNC-2D equation also yields a fair approximation for the
matrix-fluid correlations. This avenue can be further exploited
using the spatial decomposition approach\cite{Yamazaki2011} to analyze
the solvation free energy contribution in specific regions of the
adsorbate, which can be of use when dealing with functionalized
substrates. The application to three dimensional systems, mixtures and
molecular adsorbates is currently work in progress.

% If you have acknowledgments, this puts in the proper section head.
\begin{acknowledgments}

E.L and C.B.  acknowledge the support from the Direcci\'on
General de Investigaci\'on Cient\'{\i}fica  y T\'ecnica under Grant
No. FIS2010-15502. The CSIC is also
acknowledged for providing support in the form of the project PIE
201080E120. GK acknowledges financial support from the
Austrian Science Fund (FWF) under Project Nos. P23910-N16  and F41 (SFB ViCoM).
% Put your acknowledgments here.
\end{acknowledgments}

%\bibliography{ei}

\newpage

%%%%%%%%%%%%%%%%%%%%%%%%%%%%%

\end{document}